\documentstyle[amssymb,aps,eqsecnum,twocolumn]{revtex}

\newcommand{\mbb}[1]{\mbox{\boldmath $#1$}}

\begin{document}
\title{On the equivalence of the Langevin and
auxiliary field quantization methods
for absorbing dielectrics.}
\author{A.~Tip\thanks{
Electronic address: tip@amolf.nl}}
\address{FOM-Instituut voor Atoom- en Molecuulfysica\\
Kruislaan 407, Amsterdam, the Netherlands}
\author{L.~Kn\"oll, S.~Scheel\thanks{
Electronic address: scheel@tpi.uni-jena.de}, and D.-G.~Welsch}
\address{Theoretisch-Physikalisches Institut,
Friedrich-Schiller-Universit\"{a}t Jena\\
Max-Wien-Platz 1, D-07743 Jena, Germany}
\date{November 2, 2000}
\draft
\maketitle

\begin{abstract}
Recently two methods have been developed for the quantization of
the electromagnetic field in general dispersing and absorbing
linear dielectrics. The first is based upon the introduction of a
quantum Langevin current in Maxwell's equations [T.~Gruner and D.-G.~Welsch,
Phys. Rev. A {\bf 53}, 1818 (1996); Ho~Trung~Dung, L.~Kn\"{o}ll, and
D.-G.~Welsch, Phys. Rev. A{\bf \ 57}, 3931 (1998); S.~Scheel, L.~Kn\"{o}ll,
and D.-G.~Welsch, Phys. Rev. A {\bf 58}, 700 (1998)], whereas the second
makes use of a set of auxiliary fields, followed by a canonical quantization
procedure [A.~Tip, Phys. Rev. A {\bf 57}, 4818 (1998)]. We show that both
approaches are equivalent.
\end{abstract}

\narrowtext

\section{Introduction}

With the advent of modern optical materials, such as optical fibers and
photonic crystals, the problem of quantization of the electromagnetic
field in dielectrics has become an important subject and
much activity has been taking place in this field.
Quantization is required to describe the decay of embedded atoms (for
specific cases, see \cite{Polman,Ho}), the Casimir effect \cite{Lif,Mil} and
other nonclassical phenomena such as the propagation of entangled states
through dielectrics \cite{ent}. We also mention the generation of X-ray
transition radiation by fast electrons traveling through layered dielectrics
\cite{Boris}.

In the case of linear conservative dielectrics, quantization is
well-known for systems, where the permittivity (electric permeability)
$\varepsilon$ is a real constant \cite{Jauch} or a real function
of space, $\varepsilon $ $\!=$ $\!\varepsilon ({\bf r})$
\cite{KMW87,Glauber91}. Nonlinear dielectrics are discussed
in \cite{nonlin}. For dispersing and truly absorbing media the
situation is more complicated, because the permittivity is
a complex function of frequency and varies with space in general,
i.e., $\varepsilon $ $\!=$ $\!\varepsilon ({\bf r},\omega )$.
Progress in this field has been fairly recent.

Two basic approaches can be
distinguished. The first is based on the Hopfield model of a bulk dielectric
\cite{Hopfield}. The quantized electromagnetic field is coupled to a
material system described by a harmonic oscillator model, and the
Hamiltonian of the total system is diagonalized \cite{Microscopic}. A
drawback is that it becomes rather cumbersome if spatial inhomogeneities are
present \cite{Yeung}. Also the identification of the permittivity is not
trivial \cite{Gruner95,Dutra}.

The second approach starts off from the
classical phenomenological Maxwell equations, featuring a general spatially
inhomogeneous, complex, frequency-dependent permittivity
$\varepsilon ({\bf r},\omega )$ satisfying the Kramers-Kronig
relations. It has the advantage that the really measured values of the
permittivity can be used for the theoretical description of quantized
light in media. For example, in the case of
photonic crystals made up from dielectric objects (scatterers) in a
conservative, homogeneous background (such as vacuum),
$\varepsilon ({\bf r},\omega )$ is known at the outset.
Absorption is often undesirable, in particular if one is interested in
band-gap phenomena. In fact absorption prohibits the formation of
the latter \cite{Nogap}.
Band-gap photonic crystals offer many interesting technological applications
\cite{Kreta} but require a large dielectric contrast between the scatterers
and background. This can be accomplished by using small metal spheres
showing a Drude-type behavior, where the real part of the permittivity can
acquire large negative values \cite{Moroz1}. However, such systems are
always somewhat absorbing. On the positive side, absorption may be
advantageous in the case of transition radiation, where it can be used to
suppress undesired frequencies \cite{Boris}.

There have been two concepts of quantization of the phenomenological
Maxwell field for general dispersing and absorbing linear dielectrics.
The first (referred to as LN concept) is based upon the introduction of
Langevin noise current (and charge) densities,
as dictated by the fluctuation-dissipation theorem,
into the classical Maxwell equations, which
can then be transferred to quantum theory by conversion of the
electromagnetic field quantities into operators. After some earlier work
\cite{Gruner95,Loudon}, restricted to specific simple geometries, a general
formalism was put forward by some of us \cite{Gruner96,3D,QED}. In this
scheme the dyadic Green's function associated with the classical
(inhomogeneous) Helmholtz
equation plays a prominent role. Its properties come into play by deriving
the equal-time commutators for the fields, given those of the noise current
operator. In Ref.~\cite{3D} the case of a planar interface and
in Ref.~\cite{Ho,SE}
the spontaneous decay in a spherical cavity is worked out but more involved
situations can also be handled. Basically the Green's function of
the classical problem must be calculated.
For this, general methods and a variety of specific
examples are considered in Ref.~\cite{Chew}. Efficient
methods have been developed
(such as an adaptation of the KKR approach of solid state physics) for the
photonic crystal case \cite{Moroz2}.

The second concept (referred to as AF concept) developed by one
of us (AT) \cite
{Can,LAD} also starts off from the classical phenomenological
Maxwell equations. Here, the introduction of a set of
auxiliary fields (instead of a noise current) allows the replacement
of Maxwell's equations, which feature a time convolution term relating
the polarization to the electric field, by a new set of equations
for the combined set of electromagnetic and auxiliary fields but
without time convolutions. For the so extended system,
a conserved quantity, bilinear in all fields, generalizing the
electromagnetic energy, exists. Maxwell's equations are retrieved by setting
the initial auxiliary fields equal to zero. The system can then
be quantized and the conserved quantity becomes the Hamiltonian.
But now the initial auxiliary fields, being
operators, can no longer be set equal to zero.

Although the two concepts of quantization of the phenomenological
Maxwell field look quite different at first glance, they are
fully equivalent, as we show in the present paper. For this
purpose, we first review in Section \ref{sec2} the basic formulas
of the LN concept. After reviewing the AF concept, we then prove
in Section \ref{sec3} that the AF concept precisely leads to
the noise current operator in the LN concept. Some concluding
remarks are given in Section \ref{sec4}. 


\section{The Langevin noise method}
\label{sec2}

Starting point is the set of the classical macroscopic
Maxwell equations for the electromagnetic field in an
absorbing linear dielectric without free charges and currents
\begin{eqnarray}
&&
\partial_t{\bf D}({\bf r},t) =\partial_{\bf r}\times
{\bf H}({\bf r},t),
\label{2-1a}\\
&&
\partial_t{\bf B}({\bf r},t) =-\partial_{\bf r}\times
{\bf E}({\bf r},t),
\label{2-1b}\\
&&
\partial_{\bf r}\cdot {\bf D}({\bf r},t_0) = 0,
\label{2-1c}\\
&&
\partial_{\bf r}\cdot {\bf B}({\bf r},t_0) = 0,
\label{2-1d}\\
&&
{\bf D}({\bf r},t) = \varepsilon_0 {\bf E}({\bf r},t) +{\bf P}({\bf r},t),
\label{2-1e}\\
&&
{\bf P}({\bf r},t) = \varepsilon_0\int_{t_0}^t ds \,\chi ({\bf r},t-s)
{\bf E}({\bf r},s)+{\bf P}_{\rm n}({\bf r},t),
\label{2-1f}\\
&&
{\bf B}({\bf r},t) = \mu_0 {\bf H}({\bf r},t),
\label{2-1g}
\end{eqnarray}
where the initial time $t_0$ may be set to 
\mbox{$t_0$ $\!=$ $\!-\infty $}. Introducing the 
Fourier transform of the electric-field strength  according to
\begin{eqnarray}
{\bf E}({\bf r},t) &=&\int_{-\infty }^{+\infty }d\omega \exp [-i\omega t]
{\bf E}({\bf r},\omega )  \nonumber \\
&=&\int_0^{+\infty }d\omega \exp [-i\omega t]{\bf E}({\bf r},\omega )+ 
\mbox{c.c.},  \label{2-1h}
\end{eqnarray}
and the Fourier transforms of the other fields accordingly,
Eqs.~(\ref{2-1a}) -- (\ref{2-1g}) lead to
\begin{eqnarray}
&&
\partial_{\bf r} \times {\bf H}({\bf r},\omega ) = -i\omega
{\bf D}({\bf r},\omega ),
\label{2-2a}\\
&&
\partial_{\bf r} \times {\bf E}({\bf r},\omega ) = i\omega
{\bf B}({\bf r},\omega ),
\label{2-2b}\\
&&
\partial_{\bf r} \cdot {\bf D}({\bf r},\omega ) = 0,
\label{2-2c}\\
&&
\partial_{\bf r} \cdot {\bf B}({\bf r},\omega )  = 0,
\label{2-2d}\\
&&
{\bf D}({\bf r},\omega )  = \varepsilon_0 \varepsilon ({\bf r},\omega )
{\bf E}({\bf r},\omega )+{\bf P}_{\rm n}({\bf r},\omega ),
\label{2-2e}\\
&&
{\bf B}({\bf r},\omega ) = \mu_0 {\bf H}({\bf r},\omega ),
\label{2-2f}
\end{eqnarray}
where
\begin{eqnarray}
&&
\varepsilon ({\bf r},\omega )  = 1+\chi ({\bf r},\omega ),
\label{2-3a}\\
&&
\chi ({\bf r},\omega ) =\int_0^\infty  dt\exp [i\omega t]
\chi ({\bf r},t).
\label{2-3b}
\end{eqnarray}
Note that for absorbing media 
\begin{equation}
\varepsilon ({\bf r},\omega )=\varepsilon_R({\bf r},\omega )
+i\varepsilon_I({\bf r},\omega ),
\quad
\varepsilon_I({\bf r},\omega ) \ge 0.
\label{2-8}
\end{equation}

In the LN concept \cite{Gruner96,3D,QED,Buch}, Eqs.~(\ref{2-2a}) --
(\ref{2-2f}) [or Eqs.~(\ref{2-1a}) -- (\ref{2-1g})] are considered as
a set of equations for the electromagnetic field supplemented with a
noise polarization ${\bf P}_{{\rm n}}({\bf r},\omega )$ \cite{Landau}.
Its introduction arises from the necessity to fulfill the
fluctuation-dissipation theorem, because macroscopic electrodynamics
is a statistical theory. In a classical
theory the noise term can only be dropped in the zero-temperature limit, $T$
$\!\to $ $\!0$, whereas in quantum theory it is always present due to vacuum
noise.
{F}rom these arguments, the operator-valued fields (indicated with hats)
in quantum electrodynamics can be regarded as obeying
Eqs.~(\ref{2-2a}) -- (\ref{2-2f}),   
\begin{eqnarray}
&&
\partial_{\bf r}\times \hat{\bf B}({\bf r},\omega ) =
-i\frac{\omega}{c^2}\varepsilon({\bf r},\omega )
\hat{\bf E}({\bf r},\omega )+\mu_0 \hat{\bf j}_{\rm n}({\bf r},\omega ),
\label{2-5a}\\
&&
\partial_{\bf r}\times \hat{\bf E}({\bf r},\omega ) = i\omega
\hat{\bf B}({\bf r},\omega ),
\label{2-5b}\\
&&
\partial_{\bf r}\cdot \varepsilon_0\varepsilon ({\bf r},\omega )
\hat{\bf E}({\bf r},\omega ) = \hat{\rho}_{\rm n}({\bf r},\omega ),
\label{2-5c}\\
&&
\partial_{\bf r}\cdot \hat{\bf B}({\bf r},\omega ) = 0,
\label{2-5d}\\
&&
\hat{\bf D}({\bf r},\omega ) = \varepsilon_0
\varepsilon ({\bf r},\omega )\hat{\bf E}({\bf r},\omega )+
\hat{\bf P}_{\rm n}({\bf r},\omega ),
\label{2-e}
\end{eqnarray}
where $\hat{\rho}_{\rm n}({\bf r},\omega )$ and
$\hat{\bf j}_{\rm n}({\bf r},\omega )$ are the noise charge and
current densities, 
\begin{eqnarray}
&&
\hat{\rho}_{\rm n}({\bf r},\omega ) = -\partial_{\bf r}\cdot
\hat{\bf P}_{\rm n}({\bf r},\omega ),
\label{2-6a}\\
&&
\hat{\bf j}_{\rm n}({\bf r},\omega) = -i\omega
\hat{\bf P}_{\rm n}({\bf r},\omega ).
\label{2-6b}
\end{eqnarray}
Quantization is accomplished by relating the current to bosonic
vector fields according to ($\varepsilon_0\mu_0$ $\!=$ $\!c^{-2}$)
\begin{eqnarray}
\hat{\bf j}_{\rm n}({\bf r},\omega ) &=&\frac{\omega}{\mu_0c^2}
\sqrt{\frac{\hbar}{\pi \varepsilon_0}
\varepsilon_I({\bf r},\omega)} \,\hat{\bf f}({\bf r},\omega )
\nonumber \\&&
= \omega \sqrt{\frac{\hbar\varepsilon_0}{\pi} \varepsilon_I({\bf r},\omega)}
\,\hat{\bf f}({\bf r},\omega ),  \label{2-9}
\end{eqnarray}
\begin{eqnarray}
&&
\left[ \hat{\bf f}({\bf r},\omega ),
\hat{\bf f}^\dagger ({\bf r}^{\prime},\omega^\prime )\right]
= \delta ({\bf r}-{\bf r}^\prime)
\delta (\omega -\omega^\prime){\sf U},
\label{2-10a} \\
&&
\left[ \hat{\bf f}({\bf r},\omega ),
\hat{\bf f}({\bf r}^\prime,\omega^\prime)\right]
= \left[ \hat{\bf f}^\dagger ({\bf r},\omega),
\hat{\bf f}^\dagger({\bf r}^\prime ,\omega^\prime )\right] =0,
\label{2-10b}
\end{eqnarray}
where {\sf U} is the unit $3\times 3$ matrix. The fields
$\hat{\bf f}({\bf r},\omega)$ represent the fundamental
variables of the overall system. In the 
Heisenberg picture they evolve as
\begin{equation}
\hat{\bf f}({\bf r},\omega ,t)=\exp [-i\omega (t-t^\prime)] \,
\hat{\bf f}({\bf r},\omega ,t^\prime),  \label{2-11}
\end{equation}
which is governed by the Hamiltonian
\begin{equation}
\hat{H}=\int d{\bf r}\int_0^\infty d\omega \,\hbar \omega \,
\hat{\bf f}^\dagger({\bf r},\omega)
\hat{\bf f}({\bf r},\omega).  \label{2-12}
\end{equation}
The commutation relations (\ref{2-10a}) and (\ref{2-10b}) imply that
\begin{eqnarray}
\lefteqn{
\left[ \hat{\bf j}_{\rm n}({\bf r},\omega ),
\hat{\bf j}_{\rm n}^\dagger({\bf r}^\prime ,\omega^\prime )\right] }
\nonumber \\&&\quad
=\left( \frac{\omega}{\mu_0 c^2}\right)^2
\frac{\hbar }{\pi \varepsilon_0}\varepsilon_{I}({\bf r},\omega)
\delta ({\bf r}-{\bf r}^\prime) \delta (\omega -\omega^\prime){\sf U}
\nonumber \\&&\quad
=\frac{\hbar \omega^2}{\pi}\varepsilon_0
\varepsilon_{I}({\bf r},\omega) \delta ({\bf r}-{\bf r}^\prime)
\delta (\omega -\omega^\prime) {\sf U}.  \label{2-13}
\end{eqnarray}

{F}rom Eqs.~(\ref{2-5a}) and (\ref{2-5b}) it follows
that $\hat{\bf E}({\bf r},\omega )$ satisfies the equation
\begin{eqnarray}
\lefteqn{
\partial_{\bf r} \times \partial_{\bf r} \times
\hat{\bf E}({\bf r},\omega) -\frac{\omega^2}{c^2}
\varepsilon ({\bf r},\omega) \hat{\bf E}({\bf r},\omega )
}
\nonumber \\ &&\quad
= [{\sf H}_0-\frac{\omega^2}{c^2} \varepsilon({\bf r},\omega)
{\sf U}] \cdot \hat{\bf E}({\bf r},\omega )
= i\omega \mu_0 \hat{\bf j}_{\rm n}({\bf r},\omega )  \label{2-14}
\end{eqnarray}
(${\sf H}_0$ $\!=$ $\!\partial_{\bf r}\times \partial_{\bf r}$
$\!\times$ $\!=$
$\!\partial_{\bf r}\partial_{\bf r}-\partial_{\bf r}^2{\sf U}$). 
Inversion of Eq.~(\ref{2-14}) and the use of Eq.~(\ref{2-5b}) yields  
\begin{eqnarray}
\hat{\bf E}({\bf r},\omega ) 
&=&i\omega \mu_0 \int d{\bf s}\,{\sf G}({\bf r},{\bf s},\omega)
\cdot \hat{\bf j}_{\rm n}({\bf s},\omega) ,
\label{2-15a} \\
\hat{\bf B}({\bf r},\omega ) &=& (i\omega)^{-1} \partial_{\bf r} \times
\hat{\bf E}({\bf r},\omega )
\nonumber \\
&=&\mu_0\partial_{\bf r}\times \int d{\bf s}\,
{\sf G}({\bf r},{\bf s},\omega ) \cdot
\hat{\bf j}_{\rm n}({\bf s},\omega ).
\label{2-15b}
\end{eqnarray}
Here, ${\sf G}$ is the classical Green function (actually a
second-rank symmetric tensor) that satisfying the equation
\begin{equation}
\left\{ \partial_{\bf r} \partial_{\bf r}-
\left[
\partial_{\bf r}^2+
\frac{\omega^2}{c^2}\varepsilon ({\bf r},\omega )
\right]{\sf U}
\right\} \cdot
{\sf G}({\bf r},{\bf s},\omega )=\delta ({\bf r}-{\bf s}){\sf U}.
\label{2-16}
\end{equation}
Note that ${\sf G}$ corresponds to the operator
$[\partial_{\bf r}\partial_{\bf r}$ $\!-$
$\!\{\partial_{\bf r}^2$ $\!+$
$\!\omega^2\varepsilon({\bf r},\omega)/c^2\}$ $\!{\sf U}]^{-1}$,
which exists as a bounded Hilbert-space operator if
\begin{equation}
\lim_{|{\bf r}|\to\infty}\left[\varepsilon({\bf r},\omega)-1\right]
= i\,0_+,
\label{2-16a}
\end{equation}
automatically fixing the boundary conditions of ${\sf G}$ at
infinity (cf. Ref.~\cite{Nogap}).

The electric-field strength operator in the Schr\"{o}dinger
picture can then represented in the form of [cf. Eq.~(\ref{2-1h})]
\begin{equation}
\hat{\bf E}({\bf r})=\int_0^\infty d\omega \,
\hat{\bf E}({\bf r},\omega) +\mbox{H.c.},
\label{2-7}
\end{equation}
and the other field operators accordingly.
Using the properties of ${\sf G}$ it can be verified that the
standard equal-time commutation relations of quantum electrodynamics
fields are fulfilled \cite{Gruner96,3D,QED,Buch}. 
Since the latter do not depend on $\varepsilon ({\bf r},\omega )$,
the case that $\varepsilon _{I}({\bf r},\omega )$ (approximately)
vanishes in a certain $\omega $-interval can be handled by means
of a limiting procedure. It is worth noting that the
LN method has the advantage that arbitrary inhomogeneous,
anisotropic, amplifying or magnetic matter can easily be
included in the formalism \cite{Buch}.


\section{The auxiliary field method and its relation to the Langevin noise
method}
\label{sec3}


\subsection{Classical formalism}
\label{sec3a}

The AF method \cite{LAD} starts from the
zero-temperature classical Maxwell equations
(${\bf P}_{\rm n}$ $\!=$ $\!0$) and complements them
with appropriately chosen auxiliary fields. In order
to facilitate a comparison with the LN method, we shall use a setup where
only Fourier components for positive argument are used and in addition we
shall use a different gauge for the fields. We assume that
$\chi ({\bf r},t\!=\!0)$ $\!=$ $\!0$, which can be verified from
linear response theory. 
It excludes instantaneous surges at the initial time. Then, with
$\chi^\prime ({\bf r},t)$ $\!=$ $\!\partial_t \chi ({\bf r},t)$,
\begin{eqnarray}
\partial_t{\bf E}({\bf r},t) &=& c^2 \partial_{\bf r} \times
{\bf B}({\bf r},t)-\int_{-\infty}^t ds\,\chi^\prime({\bf r},t-s)
{\bf E}({\bf r},s)  \nonumber \\
&=& c^2 \partial_{\bf r}\times {\bf B}({\bf r},t)-{\bf J}({\bf r},t),
\label{3-1}
\end{eqnarray}
where ${\bf J}({\bf r},t)$ $\!=$ $\!\partial _{t}{\bf P}({\bf r},t)$ is the
polarization current density. Since $\chi ({\bf r},0)=0$ we have [the factor
2 arises from changing the range of the $\lambda $-integral from ${\Bbb R}$
in Ref.~\cite{LAD} to $[0,\infty )$]
\begin{eqnarray}
&&
\chi ({\bf r},t) = 2\int_0^{\infty }d\lambda\, \lambda ^{-1}
\sin(\lambda t) \,  \nu ({\bf r},\lambda ),
\label{3-2a}\\
&&
\chi^{\prime }({\bf r},t) = 2\int_0^{\infty }d\lambda\,
\cos(\lambda t) \,\nu({\bf r},\lambda),
\label{3-2b}
\end{eqnarray}
where $\nu ({\bf r},\lambda )$ $\!\ge$ $\!0$ for absorbing systems
considered here. Note that
\begin{equation}
\varepsilon_I({\bf r},\lambda )
=\frac{\pi}{\lambda }\, \nu ({\bf r},\lambda)
\label{3-3}
\end{equation}
($\lambda$ $\!\ge$ $\! 0$). Next we define
\begin{eqnarray}
&&
{\bf F}_1({\bf r},t) = \sqrt{\varepsilon_0}\,{\bf E}({\bf r},t),
\label{3-4b}\\
&&
{\bf F}_3({\bf r},t) = \frac{1}{\sqrt{\mu_0}}\,{\bf B}({\bf r},t)
\label{3-4c}
\end{eqnarray}
and introduce the auxiliary fields
\begin{eqnarray}
\lefteqn{
{\bf F}_2({\bf r},\lambda ,t)
= -\sqrt{\varepsilon_0}\,\sigma ({\bf r},\lambda )
}
\nonumber\\&&\hspace{6ex}
\times\,\int_{-\infty}^t ds\,\sin \lambda (t-s){\bf E}({\bf r},s),
\label{3-4d}
\end{eqnarray}
\begin{eqnarray}
\lefteqn{
{\bf F}_4({\bf r},\lambda ,t) = -\sqrt{\varepsilon_0}\,\sigma ({\bf r}
,\lambda )
}
\nonumber\\&&\hspace{6ex}
\times\,\int_{-\infty }^t ds\,\cos \lambda (t-s){\bf E}({\bf r},s),
\label{3-4e}
\end{eqnarray}
where
\begin{equation}
2\nu ({\bf r},\lambda ) = \sigma ({\bf r},\lambda )^2,
\quad \sigma ({\bf r},\lambda )\geq 0,
\label{3-4a}
\end{equation}
and note that
\begin{equation}
{\bf F}_2({\bf r},\lambda ,-\infty )={\bf F}_4({\bf r},\lambda ,-\infty)=0.
\label{3-5}
\end{equation}
It can be proved \cite{LAD} that the set of equations
\begin{eqnarray}
&&
\partial_t {\bf F}_1({\bf r},t) =
c\partial_{\bf r}\times {\bf F}_3({\bf r},t)
\nonumber\\
&&\hspace{12ex}
+\int_0^{\infty }d\lambda \sigma ({\bf r},\lambda )
{\bf F}_4({\bf r},\lambda ,t),
\label{3-6a}\\
&&
\partial_t{\bf F}_2({\bf r},\lambda ,t) =
\lambda {\bf F}_4({\bf r},\lambda ,t),
\label{3-6b}\\
&&
\partial_t{\bf F}_3({\bf r},t) = -c\partial_{\bf r}\times
{\bf F}_1({\bf r},t),
\label{3-6c}\\
&&
\partial_t {\bf F}_4({\bf r},\lambda ,t) = -\lambda
{\bf F}_2({\bf r},\lambda ,t) -\sigma ({\bf r},\lambda )
{\bf F}_1({\bf r},t)
\label{3-6d}
\end{eqnarray}
together with the initial conditions (\ref{3-5}) is
equivalent to Maxwell's equations, and the quantity
\begin{eqnarray}
\lefteqn{
{\cal E} = \textstyle\frac{1}{2}
\int d{\bf r}\,\left[{\bf E}({\bf r},t)^2+{\bf B}({\bf r},t)^2\right]
}
\nonumber \\ &&\hspace{2ex}
+ \,\textstyle\frac{1}{2}
\int d{\bf r}\int_0^\infty d\lambda\,
 \left[{\bf F}_2({\bf r}
,\lambda ,t)^2+{\bf F}_4({\bf r},\lambda ,t)^2\right]
\label{3-7}
\end{eqnarray}
is conserved in time. Note that ${\cal E}$ coincides with the
electromagnetic energy for vanishing $\chi$.

Our aim is to find a quantized
version of Eqs.~(\ref{3-6a}) -- (\ref{3-6d}). Since the
initial condition (\ref{3-5}) then loses its meaning, 
we now drop it. Setting
\begin{equation}
{\bf F}_0({\bf r},\lambda ,t)
={\bf F}_4({\bf r},\lambda ,t)-i{\bf F}_2({\bf r},\lambda ,t),
\label{3-8}
\end{equation}
we have
\begin{equation}
\partial_t{\bf F}_0({\bf r},\lambda ,t) =
-i\lambda {\bf F}_0({\bf r},\lambda ,t)
-\sigma ({\bf r},\lambda ){\bf F}_1({\bf r},t).  \label{3-9}
\end{equation}
Its solution can be written as
\begin{eqnarray}
\lefteqn{
\exp (i\lambda t) \, {\bf F}_0({\bf r},\lambda,t)
}
\nonumber \\ &&\hspace{2ex}
= {\bf F}_0^\prime({\bf r},\lambda )-\sigma ({\bf r},\lambda)
\int_{-\infty}^t ds\,\exp (i\lambda s)\,{\bf F}_1({\bf r},s).
\label{3-10}
\end{eqnarray}
Since the second term on the right hand side vanishes
as $t$ $\!\to $ $\!-\infty $, we obtain (for the limit, see the Appendix)
\begin{equation}
{\bf F}_0^\prime({\bf r},\lambda )
=\lim_{t\rightarrow -\infty }\exp(i\lambda t)\,
{\bf F}_0({\bf r},\lambda ,t).  \label{3-12}
\end{equation}
On the other hand, setting $t$ $\!=$ $\!0$ in Eq.~(\ref{3-10}) 
yields
\begin{eqnarray}
\lefteqn{
{\bf F}_0^\prime({\bf r},\lambda )
={\bf F}_0({\bf r},\lambda,0)
}
\nonumber\\&&\hspace{2ex}
+\,\sigma ({\bf r},\lambda )\int_{-\infty }^0 ds\,
\exp (i\lambda s)\,{\bf F}_1({\bf r},s).
\label{3-13}
\end{eqnarray}
With
\begin{equation}
{\bf F}_0^\prime({\bf r},\lambda ,t)
=\exp (-i\lambda t)\,{\bf F}_0^\prime({\bf r},\lambda )  \label{3-14}
\end{equation}
Eqs.~(\ref{3-6a}) -- (\ref{3-6d}) are then replaced by
\begin{eqnarray}
&&
\partial_t{\bf D}({\bf r},t) = \partial_{\bf r}\times
{\bf H}({\bf r},t),
\label{3-15a} \\
&&
\partial_t{\bf B}({\bf r},t) = -\partial_{\bf r}\times
{\bf E}({\bf r},t),
\label{3-15b} \\
&&
\partial_t{\bf F}_2^\prime({\bf r},\lambda ,t) = \lambda
{\bf F}_4^\prime({\bf r},\lambda ,t),
\label{3-15c} \\
&&
\partial_t{\bf F}_4^\prime({\bf r},\lambda ,t) = -\lambda
{\bf F}_2^\prime({\bf r},\lambda ,t),
\label{3-15d}
\end{eqnarray}
where
\begin{eqnarray}
\lefteqn{
{\bf D}({\bf r},t)=\varepsilon_0 {\bf E}({\bf r},t)
}
\nonumber \\&&\hspace{2ex}
+\,{\varepsilon_0 \int_{-\infty }^t ds\,
\chi ({\bf r},t\!-\!s)\,{\bf E}({\bf r},s)+{\bf P}^\prime({\bf r},t)}.
\label{3-15e}
\end{eqnarray}
Here,
\begin{eqnarray}
\lefteqn{
{\bf P}^\prime({\bf r},t)=\sqrt{\varepsilon_0}
\int d\lambda\, \lambda^{-1}\sigma ({\bf r},\lambda )
}
\nonumber \\&&\hspace{2ex}
\times\, \left[\cos(\lambda t)\,{\bf F}_2^\prime({\bf r},\lambda)
+\sin(\lambda t)\,{\bf F}_4^\prime({\bf r},\lambda )\right]
\label{3-15f}
\end{eqnarray}
can be regarded as being the noise polarization, with
\begin{eqnarray}
\lefteqn{
{\bf J}^\prime({\bf r},t)=\partial_t {\bf P}^\prime({\bf r},t)
=\sqrt{\varepsilon_0} \int d\lambda \, \sigma (\lambda )
}
\nonumber\\&&\hspace{4ex}
\times\, \left[\sin(\lambda t)\,{\bf F}_2^\prime({\bf r},\lambda )
-\cos(\lambda t) \,{\bf F}_4^\prime({\bf r},\lambda) \right].  \label{3-16}
\end{eqnarray}
being the associated noise current density. In Eqs.~(\ref{3-15f})
and Eqs.~(\ref{3-16}) we have set
${\bf F}_0^\prime$ $\!=$ $\!{\bf F}_4^\prime$ $\!-$
$\!i{\bf F}_2^\prime$ with ${\bf F}_2^\prime$ and
${\bf F}_4^\prime$ real.
Note that the equations of motion for the primed auxiliary fields are
decoupled from those of the electromagnetic fields.


\subsection{Hamilton formalism and quantization}
\label{sec3b}

We can interpret Eqs.~(\ref{3-15a}) -- (\ref{3-15d})
as a set of equations suitable for transferring to quantum theory.
The equations of motion for the ${\bf F}^{\prime }$-fields describe 
harmonic motions and are readily quantized, thus leading to a quantum noise
contribution in the field equations. However, there actually exists a
Hamiltonian formalism, generating the full set of field equations, which can
then be quantized \cite{Can,LAD}.
The basic equations are Eqs.~(\ref{3-6a}) -- (\ref{3-6d}),
which can be written in the compact form of 
\begin{equation}
\partial_t{\bf F}={\sf N}{\bf F}=\left(
\begin{array}{ll}
0 & {\sf N}_{em} \\
{\sf N}_{me} & 0
\end{array}
\right) \left(
\begin{array}{l}
{\bf F}_{e} \\
{\bf F}_{m}
\end{array}
\right) ,  \label{3-17}
\end{equation}
where ${\bf F}_e$ consists of ${\bf F}_1$ and ${\bf F}_2$
and ${\bf F}_m$ of ${\bf F}_3$ and ${\bf F}_4$.

In order to show the equivalence to the LN method, we
adopt a generalization of the temporal or Weyl gauge
(the T-gauge in Ref.~\cite{Can}) instead of the generalized
Coulomb gauge in Ref~\cite{LAD}. Thus we set
\begin{eqnarray}
&&
{\bf F}_e = -\partial_t \mbb{\xi}_e=-\partial_t \left(
\begin{array}{l}
\mbb{\xi}_1 \\ \mbb{\xi}_2
\end{array}
\right) ,
\label{3-18a} \\
&&
{\bf F}_m = \left(
\begin{array}{l}
{\bf F}_3 \\
{\bf F}_4
\end{array}
\right) =-{\sf N}_{me}\left(
\begin{array}{l}
\mbb{\xi}_1 \\ \mbb{\xi}_2
\end{array}
\right) ,
\label{3-18b}
\end{eqnarray}
which implies that
\begin{eqnarray}
&&
{\bf E} =-\partial_t \varepsilon_0^{-1/2} \mbb{\xi}_1,
\label{3-19a}\\
&&
{\bf B} =\partial_{\bf r}\times \varepsilon_0^{-1/2}
\mbb{\xi}_1,
\label{3-19b}\\
&&
{\bf F}_2 = -\partial_t \mbb{\xi}_2,
\label{3-19c}\\
&&
{\bf F}_4 = \sigma \mbb{\xi}_1+\lambda \mbb{\xi}_2.
\label{3-19d}
\end{eqnarray}
Obviously, ${\bf A}$ $\!=$
$\!\varepsilon_0^{-1/2}\mbb{\xi}_1$ is
the vector potential. The Lagrangian generating the equations of motion is
\begin{eqnarray}
L &=&\textstyle\frac{1}{2} \langle \partial_t \mbb{\xi}_e,
\partial_t \mbb{\xi}_e \rangle -\textstyle\frac{1}{2}
\langle {\sf N}_{me} \mbb{\xi}_e,
{\sf N}_{me} \mbb{\xi}_e \rangle  \nonumber \\
&=&\textstyle\frac{1}{2} \langle \partial_t \mbb{\xi}_e,
\partial_t \mbb{\xi}_e \rangle - \textstyle\frac{1}{2}
\langle {\sf H}_e \mbb{\xi}_e, \mbb{\xi}_e \rangle ,  \label{3-20}
\end{eqnarray}
where the (real) inner product is defined according to
\begin{eqnarray}
\langle {\bf f},{\bf g}\rangle
&=&\varepsilon_0 \int d{\bf r}\, \Big[ {\bf f}_1({\bf r})
\cdot {\bf g}_1({\bf r})
\nonumber \\
&& +\int_0^\infty d\lambda \,{\bf f}_2({\bf r},\lambda )\cdot
{\bf g}_2({\bf r},\lambda ) \Big].  \label{3-21}
\end{eqnarray}
The canonical momentum is
\begin{equation}
\mbb{\pi}_e=
{\mbb{\pi}_1({\bf r}) \choose \mbb{\pi}_2({\bf r},\lambda )}
= {\partial_t \mbb{\xi}_1 \choose \partial_tm\mbb{\xi}_2}, 
\label{3-22}
\end{equation}
thus ${\bf E}$ $\!=$ $\!-\varepsilon_0^{-1/2}\mbb{\pi}_1$, and the
Hamiltonian becomes
\begin{eqnarray}
H &=&\textstyle\frac{1}{2} \langle \mbb{\pi}_e, \mbb{\pi}_e \rangle
+\textstyle\frac{1}{2} \langle {\sf N}_{me} \mbb{\xi}_e,
{\sf N}_{me} \mbb{\xi}_e \rangle  \nonumber \\
&=&\textstyle\frac{1}{2}\langle \mbb{\pi}_e,
\mbb{\pi}_e\rangle +\textstyle\frac{1}{2}\langle {\sf H}_e
\mbb{\xi}_e,\mbb{\xi}_e\rangle .  \label{3-23}
\end{eqnarray}
Here (cf. Ref.~\cite{LAD})
\begin{equation}
{\sf H}_e = {\sf N}_{me}^\ast {\sf N}_{me} \!=\!\left(\!
\begin{array}{ll}
c^2{\sf H}_0\!+\!\chi ^{\prime }({\bf r},0) &\
\int_0^\infty d\lambda\,\lambda \sigma ({\bf r},\lambda )\cdots
\\[1ex]
\lambda \sigma ({\bf r},\lambda ) &\ \lambda^2
\end{array}
\!\right) \label{3-24}
\end{equation}
where ${\sf H}_0$ $\!=$ $\!\partial_{\bf r}\partial_{\bf r}$ $\!-$
$\!\partial_{\bf r}^2{\sf U}$ is a non-negative selfadjoint operator
in ${\cal H}_e$ $\!=$ $\!{\cal H}_1 \oplus {\cal H}_2$, ${\cal H}_1$
$\!=$ $\!L^2({\Bbb R}^3,d{\bf r};{\Bbb R}^3)$,
\mbox{${\cal H}_2$ $\!=$
$\!L^2({\Bbb R}^3\times{\Bbb R}^+,d{\bf r}d\lambda;{\Bbb R}^3)$}.
Note that the off-diagonal
terms in Eq.~(\ref{3-24}) describe the coupling of the electromagnetic and
auxiliary fields, but that there is also a shift term
$\chi^{\prime }({\bf r},0)$, and it can  be verified that
$H$ equals the conserved quantity ${\cal E}$.

Quantization is achieved by setting
\begin{eqnarray}
&&
\big[ \hat{\mbb{\xi}}_1({\bf r}),
\hat{\mbb{\pi}}_1({\bf r}^\prime )\big]
= i\hbar \delta ({\bf r}-{\bf r}^\prime ){\sf U},
\label{3-26a} \\
&&
\big[\hat{\mbb{\xi}}_2({\bf r},\lambda ),
\hat{\mbb{\pi}}_2({\bf r}^\prime ,\lambda^\prime)\big] =
i\hbar \delta ({\bf r}-{\bf r}^\prime )
\delta (\lambda -\lambda^\prime){\sf U},  \label{3-26b}
\end{eqnarray}
all other commutators being zero. The Heisenberg equations of motion
are given according to Eqs.~(\ref{3-6a}) -- (\ref{3-6d})
[or Eq.~(\ref{3-17})]. Defining $\hat{\bf F}_0({\bf r},\lambda,t)$
according to Eq.~(\ref{3-8}), it again satisfies the
(operator-valued) Eq.~(\ref{3-9}) with the solution according
to Eq.~(\ref{3-10}):
\begin{eqnarray}
\lefteqn{
\hat{\bf F}_0({\bf r},\lambda ,t) = \exp [-i\lambda t]
\hat{\bf F}_0^{\prime}({\bf r},\lambda )
}
\nonumber \\&&\hspace{2ex}
-\sigma ({\bf r},\lambda )\int_{-\infty }^t ds
\exp[-i\lambda (t-s)] \hat{\bf F}_1({\bf r},s),  \label{3-27}
\end{eqnarray}
where, according to Eq.~(\ref{3-12}) together with Eq.~(\ref{3-8}),
\begin{eqnarray}
\hat{\bf F}_0^{\prime }({\bf r},\lambda ) =
\hat{\bf F}_4^{\prime }({\bf r},\lambda ) -
i\hat{\bf F}_2^{\prime }({\bf r},\lambda ),  \label{3-28}
\end{eqnarray}
with $\hat{\bf F}_{2,4}^{\prime }({\bf r},\lambda ,t)$ and
$\hat{\bf F}_{2,4}^{\prime }({\bf r},\lambda )$
selfadjoint. Insertion of Eq.~(\ref{3-27}) in the equations of motion
results into the operator-valued Eqs.~(\ref{3-15a}) -- (\ref{3-15f}).
According to Eq.~(\ref{3-16}), the operator of the
noise current density reads
\begin{eqnarray}
\lefteqn{
\hat{\bf J}^{\prime }({\bf r},t)
=\sqrt{\varepsilon_0} \int d\lambda \, \sigma (\lambda )
}
\nonumber\\&&\hspace{4ex}
\times\, \left[\sin(\lambda t)\,\hat{\bf F}_2^\prime({\bf r},\lambda )
-\cos(\lambda t)\,\hat{\bf F}_4^\prime({\bf r},\lambda )\right].
\label{3-30} 
\end{eqnarray}
Using Eqs.~(\ref{3-26a}) and (\ref{3-26b}), we
find the (equal-time) commutation relations 
\begin{equation}
\big[ \hat{{\bf F}}_4^{\prime }({\bf r},\lambda),
\hat{{\bf F}}_2^{\prime }({\bf r}^{\prime },\lambda ^{\prime })\big]
= -i\hbar \lambda \delta ({\bf r}-{\bf  r}^{\prime })
\delta (\lambda -\lambda ^{\prime }){\sf U}  \label{3-32}
\end{equation}
and
\begin{eqnarray}
\lefteqn{
\big[ \hat{\bf J}^\prime({\bf r},\lambda ),
\hat{\bf J}^{\prime\dagger}({\bf r}^\prime ,\lambda^\prime) \big] 
}
\nonumber \\ && \hspace{2ex}
= \frac{\hslash \lambda^2}{\pi}\,
\varepsilon_0 \varepsilon_I({\bf r},\lambda )
\delta ({\bf r}-{\bf r}^{\prime })\delta (\lambda -\lambda^\prime)
{\sf U}.  \label{3-33}
\end{eqnarray}
Thus, we identify $\hat{\bf J}^\prime({\bf r},\lambda )$ with the
Langevin noise current in Section \ref{sec2},
\begin{equation}
\hat{\bf J}^\prime({\bf r},\lambda )=
\hat{\bf j}_{\rm n}({\bf r},\lambda ).  \label{3-34}
\end{equation}
With this choice, Eq.~(\ref{3-33}) exactly equals
Eq.~(\ref{2-13}). Hence, the AF formalism is 
equivalent with the LN formalism. Introducing creation and
annihilation operators according to
\begin{eqnarray}
&&
\hat{\bf F}_2^\prime({\bf r},\lambda ) = i(\hslash \lambda/2)^{1/2}
\big[
\hat{\bf b}({\bf r},\lambda )-\hat{\bf b}^\dagger({\bf r},\lambda )
\big],
\label{3-35a}\\
&&
\hat{\bf F}_4^\prime({\bf r},\lambda ) = (\hslash \lambda/2)^{1/2}
\big[
\hat{\bf b}({\bf r},\lambda )+\hat{\bf b}^\dagger({\bf r},\lambda )
\big],
\label{3-35b}
\end{eqnarray}
we have
\begin{equation}
[ \hat{\bf b}({\bf r},\lambda ),
\hat{\bf b}^\dagger({\bf r}^{\prime},\lambda^\prime)] =
\delta ({\bf r}-{\bf r}^\prime) \delta(\lambda -\lambda^\prime)
{\sf U},  \label{3-36}
\end{equation}
and hence
\begin{equation}
\hat{\bf f}({\bf r},\lambda )=-\hat{\bf b}({\bf r},\lambda )=
-(2\hslash \lambda )^{-1/2}\hat{\bf F}_0^\prime({\bf r},\lambda ).
\label{3-37}
\end{equation}


\section{Discussion}
\label{sec4}

In the LN formalism, the basic ingredient is the identification of the
(usually discarded) noise polarization
in Eq.~(\ref{2-1f}) or Eq.~(\ref {2-2e}) as the fundamental field
variable of the theory from which all properties of the
electromagnetic field can be derived by means of
Eqs.~(\ref{2-6a}), (\ref{2-6b}), (\ref{2-15a}), (\ref{2-15b}),
and Maxwell's equations (\ref{2-2a}) -- (\ref{2-2f}).
For any temperature, the fluctuation-dissipation theorem
is then satisfied, and the classical, statistical noise polarization
can be regarded as being an operator-valued quantity
in quantum theory. Then one can show
that the correct (equal-time) QED commutation relations
are satisfied.

The AF formalism starts from Maxwell's equations
without the noise polarization.
Instead, auxiliary fields
are introduced whose equations of motion eventually decouple from
those of the electromagnetic fields leaving behind a source term
in Amp\`{e}re's law. The formalism can then be cast into a Hamiltonian
form, which, upon quantization, features a noise current with the same
commutation properties as in the LN formalism.
For ease of comparison, a generalized temporal gauge
has been adopted, but the actual choice does of course
not affect the equal-time commutator
$[\hat{\bf E}({\bf r}),\hat{\bf B}({\bf r}^{\prime })]$.

It should be pointed out that Eq.~(\ref{2-14}), which plays a
crucial role in the LN concept, can be
related to the eigenvalue problem associated with ${\sf H}_e$ in
the AF formalism. Indeed, from Eqs.~(\ref{3-6a}) -- (\ref{3-6d}) we have
\begin{equation}
\partial _{t}^{2}{\bf F}_{e}(t)=-{\sf H}_{e}{\bf F}_{e}(t).  \label{4-1}
\end{equation}
Thus, in the stationary solution ${\bf F}_e(t)$ $\!=$
$\!\exp [-i\omega t]{\bf F}_e(\omega )$,
${\bf F}_e(\omega )$ solves the eigenvalue problem
\begin{equation}
{\sf H}_e {\bf F}_e(\omega )=\omega^2 {\bf F}_e(\omega ),  \label{4-2}
\end{equation}
where the first of these equations,
\begin{eqnarray}
\lefteqn{
\left[c^2 {\sf H}_0+\chi^\prime(0)\right] {\bf F}_1(\omega) }
\nonumber \\&&\hspace{2ex}
+\int_0^{\infty } d\lambda \,\lambda \sigma( \lambda)
{\bf F}_2(\lambda ,\omega ) =\omega^2{\bf F}_1(\omega ),  \label{4-3}
\end{eqnarray}
corresponds to Eq.~(\ref{2-14}), given the relation (\ref{3-12})
between ${\bf F}_0$ and $ {\bf F}_0^\prime$.
As shown in the Appendix, this relation has a
precise scattering theoretical background.

\acknowledgments

A. Tip was sponsored by the Stichting voor Fundamenteel Onderzoek
der Materie (Foundation for Fundamental Research on Matter) with
financial support from the Nederlandse Organisatie voor
Wetenschappelijk Onderzoek (Netherlands Organization for
Scientific Research).


\appendix

\section*{The relation between ${\bf F}_0$ and ${\bf F}_0^\prime$}

The existence of the limit
\begin{equation}
{\bf F}_0^\prime({\bf r},\lambda )
=\lim_{t\rightarrow -\infty }\exp(i\lambda t)\,
{\bf F}_0({\bf r},\lambda ,t)  \label{A-1}
\end{equation}
suggests that ${\bf F}_0({\bf r},\lambda ,t)$ is asymptotically free,
i.e., its motion becomes decoupled from that of the electromagnetic fields
as $t$ $\!\to $ $\!-\infty $. Such an asymptotic behavior can be
studied more precisely in terms of M{\o }ller wave operators, as we
shall now briefly discuss. We write
\begin{eqnarray}
{\sf N} &=&\left(
\begin{array}{cccc}
0 & 0 & c\partial_{\bf r}\times & \int_0^\infty
d\lambda\, \sigma ({\bf r},\lambda )\cdots \\[.5ex]
0 & 0 & 0 & \lambda \\[.5ex]\
-c\partial_{\bf r}\times & 0 & 0 & 0 \\[.5ex]
-\sigma ({\bf r},\lambda ) & -\lambda & 0 & 0
\end{array}
\right)  \nonumber \\[1ex]
&=&{\sf N}_0+{\sf N}_1,  \label{A-2}
\end{eqnarray}
where
\begin{eqnarray}
&&
{\sf N}_0 =\left(
\begin{array}{cccc}
0 & 0 & c\partial_{\bf r} \times & 0 \\[.5ex]
0 & 0 & 0 & \lambda \\[.5ex]
-c\partial_{\bf r} \times & 0 & 0 & 0 \\[.5ex]
0 & -\lambda & 0 & 0
\end{array}
\right) ,\quad
\label{A-3a} \\[1ex]
&&
{\sf N}_{1} =\left(
\begin{array}{cccc}
0 & 0 & 0 & \int_0^\infty d\lambda\,
\sigma ({\bf r},\lambda ) \cdots \\[.5ex]
0 & 0 & 0 & 0 \\[.5ex]
0 & 0 & 0 & 0 \\[.5ex]
-\sigma ({\bf r},\lambda ) & 0 & 0 & 0
\end{array}
\right) .  \label{A-3b}
\end{eqnarray}
Let
\begin{equation}
{\sf P}_{\rm aux}=\left(
\begin{array}{llll}
0 & 0 & 0 & 0 \\
0 & 1 & 0 & 0 \\
0 & 0 & 0 & 0 \\
0 & 0 & 0 & 1
\end{array}
\right)  \label{A-4}
\end{equation}
be the projector upon the auxiliary fields. Then it can be shown by standard
methods that the M{\o}ller operators
\begin{equation}
\Omega_\pm =\lim_{t\rightarrow \pm \infty }
\exp (-{\sf N}t)\,\exp ({\sf N}_0 t)\,{\sf P}_{\rm aux}  \label{A-5}
\end{equation}
exist in the strong sense (for the underlying Hilbert space and other
details, cf. \cite{LAD}). A quite more elaborate analysis shows that the
adjoints $\Omega_\pm^\ast$ can also be given as strong limits, i.e.,
\begin{equation}
\Omega_\pm^\ast=\lim_{t\rightarrow \pm \infty }{\sf P}_{\rm aux}
\exp(-{\sf N}_0 t)\,\exp({\sf N}t)  \label{A-6}
\end{equation}
exist, implying that
\begin{equation}
{\sf P}_{\rm aux}{\bf F}(t)\stackrel{t\rightarrow \pm \infty }{\thicksim }
\exp({\sf N}_0 t)\,\Omega_\pm^\ast {\bf F}(0),  \label{A-7}
\end{equation}
i.e. the motion of the auxiliary fields becomes decoupled from that of the
electromagnetic ones for large times. Physically, this can be understood by
observing that for a finite dielectric the auxiliary fields do not propagate
(they are confined to the dielectric), whereas the electromagnetic fields
propagate away. This gives a rigorous underpinning of the existence of the
limits in Eq.~(\ref{3-12}). We can arrive at Eq.~(\ref{3-13}),
starting from Eq.~(\ref{A-7}), noting that
\begin{eqnarray}
\lefteqn{
\Omega_\pm^\ast{\bf F}(0)={\sf P}_{\rm aux}{\bf F}(0)
}
\nonumber \\&&\hspace{2ex}
+\int_0^{\pm \infty} dt\,{\sf P}_{\rm aux}
\exp (-{\sf N}_0 t)\,({\sf N}\!-\!{\sf N}_0)\,\exp ({\sf N}t)\,{\bf F}(0)
\nonumber \\ &&\hspace{2ex}
={\sf P}_{\rm aux}{\bf F}(0)+\int_0^{\pm \infty }dt\,{\sf P}_{\rm aux}
\exp (-{\sf N}_0 t)\,{\sf N}_1 {\bf F}(t),  \label{A-8}
\end{eqnarray}
and working things out, which leads to
\begin{eqnarray}
\lefteqn{
{{\bf F}_2^\prime  \choose {\bf F}_4^\prime}
=
{(\Omega_-^\ast {\bf F})_2 \choose (\Omega_-^\ast {\bf F})_4}
=
{{\bf F}_2(0) \choose {\bf F}_4(0)}
}
\nonumber \\&&\hspace{2ex}
+\int_{-\infty }^0ds\,
{-\sin( \lambda s) \choose \hspace{2ex} \cos( \lambda s)}
\sigma (\lambda )\,{\bf F}_1(s).  \label{A-9}
\end{eqnarray}

Following the procedure given above, one can show an equivalent
expression to Eq.~(\ref{3-13}) valid in the quantum case.


\end{document}